# Security Flaws in a Recent Ultralightweight RFID Protocol

Pedro Peris-Lopez *    Julio C. Hernandez-Castro
Juan M. E. Tapiador    Jan C.A. van der Lubbe

August 29, 2018

**Abstract**

In 2006, Peris-Lopez et al. [1, 2, 3] initiated the design of ultralightweight RFID protocols –with the UMAP family of protocols– involving only simple bitwise logical or arithmetic operations such as bitwise XOR, OR, AND, and addition. This combination of operations was revealed later to be insufficient for security. Then, Chien et al. proposed the SASI protocol [4] with the aim of offering better security, by adding the bitwise rotation to the set of supported operations. The SASI protocol represented a milestone in the design of ultralightweight protocols, although certain attacks have been published against this scheme [5, 6, 7]. In 2008, a new protocol, named Gossamer [8], was proposed that can be considered a further development of both the UMAP family and SASI. Although no attacks have been published against Gossamer, Lee et al. [9] have recently published an alternative scheme that is highly reminiscent of SASI. In this paper, we show that Lee et al.'s scheme fails short of many of its security objectives, being vulnerable to several important attacks like traceability, full disclosure, cloning and desynchronization.
**Keywords:** RFID, Authentication, Ultralightweight protocols, Cryptanalysis

## 1 Introduction

In an RFID system, objects are labelled with a tag. Each tag contains a microchip with a certain (generally limited) amount of computational and storage capabilities, and a coupling element. Such devices can be classified according to their memory type and power source. Another relevant parameter is tag price, which creates a broad distinction between high-cost and low-cost RFID tags.

*Corresponding Author: Delft University of Technology (TU-Delft), Faculty of Electrical Engineering, Mathematics, and Computer Science (EEMCS), Information and Communication Theory group (ICT). P.O. Box 5031 2600 GA, Delft, The Netherlands; E-mail: P.PerisLopez@tudelft.nl.

The *rule of thumb* of gate cost says that every extra 1,000 gates increases chip price by 1 cent [10].

In [4], Chien proposed a tag classification mainly based on which were the operations supported on-chip. High-cost tags are divided into two classes: "full-fledged" and "simple". Full-fledged tags support on-board conventional cryptography like symmetric encryption, cryptographic one-way functions and even public key cryptography. Simple tags can support random number generators and one-way hash functions. Likewise, there are two classes for low-cost RFID tags. "Lightweight" tags are those whose chip supports a random number generation and simple functions like a Cyclic Redundancy Checksum (CRC), but not cryptographic hash function. "Ultralightweight" tags can only compute simple bitwise operations like XOR, AND, OR, etc.

In this paper we focus in the latter category of ultralightweight tags. These tags represent the greatest challenge in terms of security, due to their expected wide deployment and, at the same time, extremely limited capabilities.

## 2  Related Work

In 2006, Peris et al. proposed a family of Ultralightweight Mutual Authentication Protocols (henceforth referred to as the UMAP family of protocols). Chronologically, $M^2AP$ [1] was the first proposal, followed by EMAP [2] and LMAP [3]. Although some vulnerabilities were discovered (active attacks [11, 12] and later on passive attacks [13, 14]) which rendered those first proposals insecure, they were an interesting advance in the field of lightweight cryptography for low-cost RFID tags.

In 2007, Hung-Yu Chien published a very interesting ultralightweight authentication protocol providing Strong Authentication and Strong Integrity (SASI) for very low-cost RFID tags [4]. The SASI protocol is highly reminiscent of the UMAP family, and more concretely, of the LMAP protocol. The main difference between these two protocols is the inclusion of the rotation in the set of operations supported by each tag. Indeed, the messages transmitted in to the insecure channel in UMAP family are computed by the composition of triangular-functions (e.g. addition modulo 2 or bitwise OR, etc.) -easily implemented in hardware- which finally results in another triangular-function [15]. A triangular-function has the property that output bits only depend of the leftmost input bits, instead of all input bits. This undesirable characteristic (lack of diffusion) greatly facilitated the analysis of the messages transmitted in the UMAP family of protocols, and thus the work of the cryptanalyst.

SASI represented a considerable advance towards the design of a secure ultralightweight protocol. However, certain important attacks have been published. First, Sun et al. proposed two desynchronization attacks. In [6], it was proposed a denial-of-service and traceability attack. Then, D'Arco et al. [7] proposed another desynchronization attack and an identity disclosure attack. In [16], Phan shows how a passive attacker can track tags, violating the location privacy of tags' holder. Finally, Hernandez-Castro et al. [17] recently proposed a full dis-closure attack, but the authors assume modular rotation instead of hamming weight rotation.

In 2008, the Gossamer protocol [8] was proposed as a further development upon both the UMAP family and the SASI protocol. So far, this scheme seems the most secure ultralightweight authentication protocol for low-cost RFID tags, as no attacks have been published. As alternative to Gossamer, Lee et al. recently published a new ultralightweight RFID Protocol with mutual authentication (UMA-RFID in in the following) [9]. The analysis of this recent protocol is the subject of this paper.

## 3 Lee et al. Ultralightweight RFID Protocol with Mutual Authentication

Tag, reader and back-end database are the three entities involved in the protocol. Each tag has a static identifier ($ID$). A pseudonym -dynamic temporary identifier- ($IDT$) and a secret key ($K$) are shared between the tag and the reader. Indeed, the old and the potential new values of the pair $\{IDT, K\}$ are both kept in the tag to hinder desynchronization attacks. The length of the variables is 128 bits. The channel between the tag and the reader is insecure due to the open nature of the radio channel. In contrast, a secure channel is assumed for the communications between the reader and the back-end database.

Tags are limited to bitwise operations (i.e. bitwise XOR, OR and AND) and left bitwise rotation. Specifically, $Rot(A, B)$ symbolizes that the vector $A$ is subjected to a circular shift of $n$ bit positions, where $n$ is the hamming weight of vector $B$ (i.e. $n = hw(B)$). Readers are limited to the same set of operations and have the extra capability of random number generation.

We described the messages exchanged in the protocol below (see also Figure 1). First, the reader ($\mathcal{R}$) and the tag ($\mathcal{T}$) are mutually authenticated (authentication phase). Then, the reader and the tag, respectively, update their private information $\{IDT, K\}$ shared and kept synchronized between them (updating phase).

1. **Authentication Phase**

$\mathcal{T} \to \mathcal{R} : IDT_i$ In the session $i$-th, the reader sends a request message to the tag. Then, the tag backscatters its pseudonym ($IDT_i$) to provide anonymous identification.

$\mathcal{R} \to \mathcal{T} : A_i, B_i$ Upon receiving $IDT_i$, the reader looks up in the database the secret key associated to $\mathcal{T}$. Then, it generates a new random value $N_i$ and computes the authentication messages $A_i$ and $B_i$:

$$A_i = K_i \oplus N_i \qquad (1)$$
$$B_i = Rot(K_i, K_i) \oplus Rot(N_i, N_i) \qquad (2)$$

The reader sends $\{A_i, B_i\}$ to the tag.

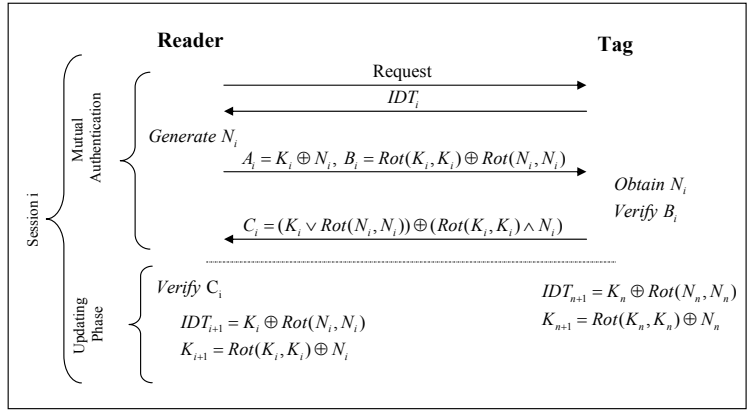

Figure 1: Ultralightweight RFID Protocol with Mutual Authentication

$\mathcal{T} \to \mathcal{R} : C_i$ After receiving $\{A_i, B_i\}$, the tag obtains $N'_i$ from message $A_i$ ($N'_i = A_i \oplus K_i$) and computes its local version of $B_i$ ($B'_i = Rot(K_i, K_i) \oplus Rot(N_i, N_i)$). If $B_i = B'_i$, the reader is authenticated. Then, the tag computes the authentication message $C_i$:

$$C_i = (K_i \vee Rot(N_i, N_i)) \oplus ((Rot(K_i, K_i) \wedge N_i) \qquad (3)$$

Finally, the tag sends $C_i$ to the reader.

$\mathcal{R}$: Upon receiving $C_i$, the reader checks its correctness to authenticate the tag.

2. **Updating Phase** Upon the reader authentication (messages $A_i$, $B_i$), the tag updates its secret information when message $C_i$ is sent. The updating in the reader is conditioned to the valid authentication of the tag (message $C_i$). Specifically, the updating phase is defined by the equations below:

$$IDT_{i+1} = K_i \oplus Rot(N_i, N_i) \qquad (4)$$
$$K_{i+1} = Rot(K_i, K_i) \oplus N_i \qquad (5)$$

# 4 Security Analysis

In this section, we describe the security vulnerabilities of Lee et al. protocol.

## 4.1 Traceability Attack

Traceability is one of the most important security threats linked to RFID technology. Location privacy is compromised when tags answer readers queries with a constant, static value, something that curiously happens in numerous commercial tags. An encrypted version of the static identifier may be used for privacy protection, but an attacker could still track the tag's holder as the tag keeps on sending a constant value. So it seems necessary to anonymize tags' answers by the inclusion of nonces. However, the simple use of random numbers by itself does not guarantee that a protocol will be resistant to traceability attacks [18].

The traceability problem has attracted a lot of research. In [19], Juels and Weis give a formal definition of the untraceability model. The same definition, though with a style more similar to that used for security protocols, is described by Phan in his attack against the SASI protocol [16].

In RFID schemes, tags $\mathcal{T}$ and readers $\mathcal{R}$ interact in protocol sessions. In general terms, the adversary ($\mathcal{A}$) controls the communications between all the participants and interacts passively or actively with them. Specifically, $\mathcal{A}$ can run the following queries:

- Execute($\mathcal{R}$, $\mathcal{T}$, $i$) query. This models a passive attacker. $\mathcal{A}$ eavesdrops on the channel, and gets read access to the exchange of messages between $\mathcal{R}$ and $\mathcal{T}$ in session $i$ of a genuine protocol execution.

- Send($\mathcal{X}$, $\mathcal{Y}$, $M$, $i$) query. This models that the message $M$ sends from $\mathcal{X}$ to $\mathcal{Y}$ in session $i$ is blocked or altered (e.g. flipping one bit), preventing its correct reception.

- Test($i$, $\mathcal{T}_0$, $\mathcal{T}_1$) query. This does not model any ability of $\mathcal{A}$, but it is necessary to define the untraceability test. When this query is invoked for session $i$, a random bit is generated $b \in \{0,1\}$. Then, the pseudonym $IDT_i^{\mathcal{T}_i}$ from the set $\{IDT_i^{\mathcal{T}_0}, IDT_i^{\mathcal{T}_1}\}$ and corresponding to tags $\{\mathcal{T}_0, \mathcal{T}_1\}$ is given to $\mathcal{A}$.

Upon definition of the adversary's abilities, the untraceability problem can be defined as a game $\mathcal{G}$ divided into the following phases:

**Phase 1 (Learning):** $\mathcal{A}$ can send Execute and Send queries. So, $\mathcal{A}$ eavesdrops messages -passive attack- passed over the channel and have the ability of blocking -active attack- certain messages.

**Phase 2 (Challenge):** $\mathcal{A}$ chooses two fresh tags whose associated identifiers are $ID_0$ and $ID_1$. Then he sends Test($i$, $\mathcal{T}_0$, $\mathcal{T}_1$) query. As result, $\mathcal{A}$ is given a dynamic temporary identifier $IDT_i^{\mathcal{T}_i}$ from the set $\{IDT_i^{\mathcal{T}_0}, IDT_i^{\mathcal{T}_1}\}$, which depends on a chosen random bit $b \in \{0,1\}$.

**Phase 3 (Guessing)** $\mathcal{A}$ finishes the game and outputs a bit $d$ ($d \in \{0,1\}$) as its conjecture of the value of $b$.

$\mathcal{A}$'s success in winning $\mathcal{G}$ is equivalent to the success of breaking the untraceability property offered by the protocol. So the advantage of $\mathcal{A}$ in distinguishing whether the messages correspond to $\mathcal{T}_0$ or $\mathcal{T}_1$, is defined as below:

$$Adv_{\mathcal{A}}^{UNT}(t, r_1, r_2) = |Pr[d=b] - \frac{1}{2}| \qquad (6)$$

where $t$ is a security parameter (i.e. the bit length of the key shared by the tag and the reader) and $r_1$ and $r_2$ are the number of times $\mathcal{A}$ can run Execute and Send queries respectively.

**Definition** An RFID protocol in an RFID system (S= $\{R_i, \mathcal{T}_0, \mathcal{T}_1, ....\}$) in which an adversary $\mathcal{A}$ can invoke {Execute($\mathcal{R}$, $\mathcal{T}$, $i$), Send($\mathcal{X}$, $\mathcal{Y}$, $M$,$i$), Test( $i$, $\mathcal{T}_0$, $\mathcal{T}_1$)} in a game $\mathcal{G}$, offers resistance against traceability if:

$$Adv_{\mathcal{A}}^{UNT}(t, r_1, r_2) < \varepsilon(t, r_1, r_2) \qquad (7)$$

$\varepsilon(.)$ being some negligible function.

We will show how the UMA-RFID scheme does not guarantee privacy location, being possible to track tags.

**Theorem 4.1** *The UMA-RFID protocol, on an RFID system (S= $\{R_i, \mathcal{T}_0, \mathcal{T}_1, ....\}$) in which an adversary $\mathcal{A}$ can invoke two* Execute($\mathcal{R}$, $\mathcal{T}$, $i$) *queries, one* Send($\mathcal{X}$, $\mathcal{Y}$, $M$,$i$) *query, and, subsequently, one* Test( $i$, $\mathcal{T}_0$, $\mathcal{T}_1$) *query in an untraceability game $\mathcal{G}$, is vulnerable to traceability attacks, since the advantage for an adversary to win $\mathcal{G}$ is significant:* $Adv_{\mathcal{A}}^{UNT}(t, 2, 1) = 0.5 \gg \varepsilon(t, 2, 1)$.

**Proof** Specifically, an adversary $\mathcal{A}$ performs the following steps:

**Phase 1 (Learning):** $\mathcal{A}$ sends two $\{\text{Execute}(\mathcal{R}, \mathcal{T}_0, i)\}_{i=n}^{n+1}$ queries and one Send($\mathcal{T}_0$, $\mathcal{R}$, $C_{n+1}^{\mathcal{T}_0}$, $n$+1) query to $\mathcal{T}_0$. $\mathcal{A}$ acquires the tuple $\{IDT_n^{\mathcal{T}_0}, A_n^{\mathcal{T}_0}, B_n^{\mathcal{T}_0}, C_n^{\mathcal{T}_0}\}$ and $\{IDT_{n+1}^{\mathcal{T}_0}\}$ where

$$B_n^{\mathcal{T}_0} = Rot(K_n^{\mathcal{T}_0}, K_n^{\mathcal{T}_0}) \oplus Rot(N_n^{\mathcal{T}_0}, N_n^{\mathcal{T}_0}) \qquad (8)$$
$$IDT_{n+1}^{\mathcal{T}_0} = K_n^{\mathcal{T}_0} \oplus Rot(N_n^{\mathcal{T}_0}, N_n^{\mathcal{T}_0}) \qquad (9)$$

Send($\mathcal{T}$, $\mathcal{R}$, $C_{n+1}^{\mathcal{T}_0}$, $n$+1) query frustrates the correct reception of message $C_{n+1}^{\mathcal{T}_0}$ in session $n+1$, which avoids the updating of the secret key and the pseudonym in the reader. The tag is thus identified using the old pair $\{IDT_{n+1}^{\mathcal{T}_0}, K_{n+1}^{\mathcal{T}_0}\}$ in the next session $n+2$.

**Phase 2 (Challenge):** $\mathcal{A}$ chooses two fresh tags whose associated identifiers are $ID_0$ and $ID_1$. Then he sends a Test($n+2$, $\mathcal{T}_0$, $\mathcal{T}_1$) query. As result, $\mathcal{A}$ is given a dynamic temporary identifier $IDT_{n+2}^{\mathcal{T}_i}$ from the set $\{IDT_{n+2}^{\mathcal{T}_0}, IDT_{n+2}^{\mathcal{T}_1}\}$, which depends on a chosen random bit $b \in \{0, 1\}$.

**Phase 3 (Guessing)** $\mathcal{A}$ finishes $\mathcal{G}$ and outputs a bit $d$ ($d \in \{0, 1\}$) as its conjecture of the value $b$. In particular, we propose the following procedure to obtain value $d$:

1. From Equation (8) and (9), the following constant value associated with $\mathcal{T}_0$ is obtained by the adversary:

$$X = B_n^{\mathcal{T}_0} \oplus IDT_{n+1}^{\mathcal{T}_0} = Rot(K_n^{\mathcal{T}_0}, K_n^{\mathcal{T}_0}) \oplus K_n^{\mathcal{T}_0} \qquad (10)$$

2. $\mathcal{A}$ calculates the XOR between the $B_n^{\mathcal{T}_0}$ value captured in the learning phase (Equation (8)) and the pseudonym presented in the challenge phase:

$$Y = \begin{cases} B_n^{\mathcal{T}_0} \oplus IDT_{n+2}^{\mathcal{T}_0} = B_n^{\mathcal{T}_0} \oplus IDT_{n+1}^{\mathcal{T}_0} = X & \text{if } b = 0 \\ B_n^{\mathcal{T}_0} \oplus IDT_{n+2}^{\mathcal{T}_1} \neq X & \text{if } b = 1 \end{cases} \qquad (11)$$

3. $\mathcal{A}$ utilizes the following simple decision rule:

$$d = \begin{cases} \text{if } X = Y & d = 0 \\ \text{if } X \neq Y & d = 1 \end{cases} \quad (12)$$

So the use of random numbers does not prevent in this case the attacker from associating the tags's answers with its holder, with a 100% probability of success.

∎

## 4.2 Full Disclosure, Cloning, and Desynchronization Attacks

The tag and the reader share a secret key. The main purpose of this key is the authentication of both entities. The key is combined with a random number to hamper its acquisition by the attacker when passed over the insecure channel. The above idea is well conceived but the protocol abuses of the usage of $Rot(K_i, K_i)$ and $Rot(N_i, N_i)$. Indeed, this fact facilitates a sort of linear cryptanalysis of the scheme, despite of the combination of triangular and non-triangular functions.

**Theorem 4.2** *In the UMA-RFID protocol, a passive attacker, after eavesdropping two consecutive authentication sessions $\{n, n+1\}$ between an authentic tag ($\mathcal{T}$) and a genuine reader ($\mathcal{R}$), can discover the secret key shared by these two entities by simply computing an XOR among some of the public messages transmitted over the radio channel:*

$$K_{n+1} = A_n \oplus B_n \oplus IDT_{n+1} \quad (13)$$

**Proof** We start describing the messages exchanged in sessions $\{n, n+1\}$:

**Session n:** $\{IDT_n, A_n, B_n, C_n\}$ where

$$A_n = K_n \oplus N_n \quad (14)$$
$$B_n = Rot(K_n, K_n) \oplus Rot(N_n, N_n) \quad (15)$$

**Session n + 1:** $\{IDT_{n+1}, A_{n+1}, B_{n+1}, C_{n+1}\}$ where

$$IDT_{n+1} = K_n \oplus Rot(N_n, N_n) \quad (16)$$
$$A_{n+1} = K_{n+1} \oplus N_{n+1} \quad (17)$$
$$B_{n+1} = Rot(K_{n+1}, K_{n+1}) \oplus Rot(N_{n+1}, N_{n+1}) \quad (18)$$
$$C_{n+1} = (K_{n+1} \vee Rot(N_{n+1}, N_{n+1})) \quad (19)$$
$$\oplus (Rot(K_{n+1}, K_{n+1}) \wedge N_{n+1}) \quad (20)$$

The secret key of the tag in session $n+1$ is described by the equation below:

$$K_{n+1} = Rot(K_n, K_n) \oplus N_n \quad (21)$$

Finally, the attacker can acquire the actual secret key ($K_{n+1}$) of the tag by computing the XOR between the public messages $A_n$, $B_n$ and $IDT_{n+1}$ (see Equations (14), (15) and (16)):

$$\begin{aligned}
A_n \oplus B_n \oplus IDT_{n+1} &= \\
&= K_n \oplus N_n \oplus Rot(K_n, K_n) \oplus Rot(N_n, N_n) \oplus K_n \oplus Rot(N_n, N_n) \\
&= (K_n \oplus K_n) \oplus N_n \oplus Rot(K_n, K_n) \oplus (Rot(N_n, N_n) \oplus Rot(N_n, N_n)) \\
&= (0x0) \oplus N_n \oplus Rot(K_n, K_n) \oplus (0x0) \\
&= N_n \oplus Rot(K_n, K_n) = Rot(K_n, K_n) \oplus N_n = K_{n+1} \qquad (22)
\end{aligned}$$

∎

RFID tags are usually not designed to be tamper resistant, because this will significantly increase their price. An active attacker may tamper with the tag in order to read from or write to its memory in which secret values are stored. Low-cost RFID tags cannot offer protection to this sort of attacks but should be resistant, at least, to passive attacks. We show now how a passive attacker is able to clone a tag after revealing the whole secrets stored in the tag, but without requiring any physical manipulation of it.

**Theorem 4.3** *In the UMA-RFID protocol, a passive attacker, after eavesdropping two consecutive authentication sessions $\{n, n+1\}$ between an authentic tag ($\mathcal{T}$) and a genuine reader ($\mathcal{R}$), can clone the tag by computing:*

$$\begin{aligned}
IDT_{n+2} &= K_{n+1} \oplus Rot(N_{n+1}, N_{n+1}) & (23) \\
K_{n+2} &= Rot(K_{n+1}, K_{n+1}) \oplus N_{n+1} & (24)
\end{aligned}$$

**Proof** From Theorem 2, an adversary can discover the actual secret key of the tag ($K_{n+1}$) after eavesdropping messages $\{IDT_n, A_n, B_n, C_n\}$ exchanged in session $n$ and the dynamic temporary identifier ($IDT_{n+1}$) in session $n+1$.

$$K'_{n+1} = A_n \oplus B_n \oplus IDT_{n+1} \qquad (25)$$

Then, the adversary can obtain the random number associated to the session $n+1$ by computing an XOR between the message $A_{n+1}$ and the key $K_{n+1}$. Then, message $B$ can be used to check its correctness.

$$\begin{aligned}
N'_{n+1} &= K'_{n+1} \oplus A_{n+1} & (26) \\
B_{n+1} &\stackrel{?}{=} Rot(K'_{n+1}, K'_{n+1}) \oplus Rot(N'_{n+1}, N'_{n+1}) & (27)
\end{aligned}$$

Once the actual key ($K_{n+1}$) and the random number ($N_{n+1}$) linked to session $n+1$ are known by the attacker, the new state can be computed by using these values:

$$\begin{aligned}
IDT_{n+2} &= K'_{n+1} \oplus Rot(N'_{n+1}, N'_{n+1}) & (28) \\
K_{n+2} &= Rot(K'_{n+1}, K'_{n+1}) \oplus N'_{n+1} & (29)
\end{aligned}$$

Finally, the attacker can copy the above values to the memory of a blank tag, which results in a cloning attack (having an undistinguishable copy of an authentic tag). ∎

Tags and readers have to remain synchronized to run the protocol successfully. The authors take the extra precaution of storing the old and potential new values of the pair $\{IDT, K\}$ to fight against desynchronization attacks, but in this case this well-known approach in the literature is not enough. Despite of this countermeasure, an attacker is able to desynchronize a tag and a reader exploiting Theorem 2.

**Theorem 4.4** *In the UMA-RFID protocol, a passive attacker, after eavesdropping two consecutive authentication sessions $\{n, n+1\}$ and performing a man-in-the-middle attack between an authentic tag ($\mathcal{T}$) and a genuine reader ($\mathcal{R}$), can desynchronize these two entities by sending:*

$$A_{n+1} = K_{n+1} \oplus N^*_{n+1} \tag{30}$$
$$B_{n+1} = Rot(K_{n+1}, K_{n+1}) \oplus Rot(N^*_{n+1}, N^*_{n+1}) \tag{31}$$
$$C_{n+1} = (K_{n+1} \vee Rot(N_{n+1}, N_{n+1})) \oplus (Rot(K_{n+1}, K_{n+1}) \wedge N_{n+1}) \tag{32}$$

**Proof** Taking advantage of Theorem 2 any adversary, after eavesdropping messages $\{IDT_n, A_n, B_n, C_n\}$ exchanged in session $n$ and the dynamic temporary identifier ($IDT_{n+1}$) of session $n+1$, gets the actual secret key of the tag ($K_{n+1}$).

$$K'_{n+1} = A_n \oplus B_n \oplus IDT_{n+1} \tag{33}$$

Then, the attacker starts the man-in-the-middle attack. Specifically, the attacker intercepts messages $\{A_{n+1}, B_{n+1}\}$ (see Equations (17) and (18)) and sends $\{A^*_{n+1}, B^*_{n+1}\}$ linked to the random number $N^*_{i+1}$:

$$A^*_{n+1} = K'_{n+1} \oplus N^*_{n+1} \tag{34}$$
$$B^*_{n+1} = Rot(K'_{n+1}, K'_{n+1}) \oplus Rot(N^*_{n+1}, N^*_{n+1}) \tag{35}$$

Finally, the attacker intercepts the answer $C^*_{n+1}$ of the tag, and computes the answer $C'_{n+1}$ to the original messages $\{A_{n+1}, B_{n+1}\}$ sent by the genuine reader:

$$N'_{n+1} = K'_{n+1} \oplus A_{n+1} \tag{36}$$
$$B_{n+1} \stackrel{?}{=} Rot(K'_{n+1}, K'_{n+1}) \oplus Rot(N'_{n+1}, N'_{n+1}) \tag{37}$$
$$C'_{n+1} = (K'_{n+1} \vee Rot(N'_{n+1}, N'_{n+1})) \oplus (Rot(K'_{n+1}, K'_{n+1}) \wedge N'_{n+1}) \tag{38}$$

After the mutual authentication between the tag and the reader, both entities update their internal secret values:

| Tag | Reader |
|---|---|
| $IDTN^*_{n+2} = K'_{n+1} \oplus Rot(N^*_{n+1}, N^*_{n+1})$ | $IDT'_{n+2} = K'_{n+1} \oplus Rot(N'_{n+1}, N'_{n+1})$ |
| $K^*_{n+2} = Rot(K'_{n+1}, K'_{n+1}) \oplus N^*_{n+1}$ | $K'_{n+2} = Rot(K'_{n+1}, K'_{n+1}) \oplus N'_{n+1}$ |

So the adversary deceives the tag and the reader into thinking that the random number associated to the session $n + 1$ is $N^*_{n+1}$ or $N'_{n+1}$ respectively. As a consequence of this fact, the tag a the reader lose their synchronization after the completion of the updating phase. ∎

To further clarify the attacks previously described, Figures 2(a) and 2(b) illustrate the messages exchanged.

As an easy alternatively to the last attack presented, an adversary can desynchronize tags and readers using the non-resistance of bitwise operations to active attacks [20]. The adversary can reuse old values, transmitted in the channel, to compute new valid authentication messages. Specifically, an XOR operation between the captured value and a constant value properly selected (e.g. $A_{i+1} = A_i \oplus 0x0005$) is enough to achieve this objective.

**Theorem 4.5** *In the UMA-RFID protocol, a passive attacker, after eavesdropping an authentication session n between an authentic tag ($\mathcal{T}$) and a genuine reader ($\mathcal{R}$), can desynchronize these two entities by sending: $A_{n+1} = A_n \oplus C_1$, $B_{n+1} = B_n \oplus C_2$, where $\{C_i\}_{i=1}^2$ are any constant values whose hamming weight is exactly 2.*

**Proof** First, the reader eavesdrops the messages $\{IDT_n, A_n, B_n, C_n\}$ passed over the channel in session $n$ where

$$A_n = K_n \oplus N_n \tag{39}$$
$$B_n = Rot(K_n, K_n) \oplus Rot(N_n, N_n) \tag{40}$$

After the mutual authentication, the tag and the reader update their secret values $\{IDT_{n+1}, K_{n+1}\}$. Indeed the tag stores the old and the potential new values with the objective of preventing desynchronization attacks. However, the adversary may exploit this fact -simulating the incorrect reception of $C$ message and using the old values in a new authentication- provoking a new updating in the tag but not in the reader. Specifically, the adversary follows the experiment described below:

1. **Initialization.** The adversary randomly selects the $C_1$ value, with the restriction that its hamming weigh is 2 (i.e. $hw(C_1) = 2$).

2.0. **Selection of the mask.**] The adversary picks up a $C_2$ value from the subset of $x \in \{0, 1, ..., 2^L\}$ that satisfies $hw(x) = 2$, where $L$ is the length of the variables used (i.e. $n = 128$ in Lee et al. protocol [9]).

2.1 **Authentication.** The adversary computes and sends to the legitimate tag the authentication messages:

$$A_{n+1} = A_n \oplus C_1 = K_n \oplus N_n \oplus C_1 \tag{41}$$
$$B_{n+1} = B_n \oplus C_2 = Rot(K_n, K_n) \oplus Rot(N_n, N_n) \oplus C_2 \tag{42}$$

2.2 **Check of $C_2$.** If the tag accepts $\{A_{n+1}, B_{n+1}\}$ and replies $C_{n+1}$ to the adversary, it proves the success of the attack launched. Otherwise, the process is repeated from Step 2.0.

Table 1: Performance Comparison of Ultralightweight Authentication Protocols

|  | UMAP family [1, 2, 3] | SASI [4] | UMA-RFID [9] | Gossamer [8] |
|---|---|---|---|---|
| Resistance to Desynchronization Attacks | No | No | No | Yes |
| Resistance to Disclosure Attacks | No | No | No | Yes |
| Privacy and Anonymity | No | No | No | Yes |
| Mutual Auth. and Forward Security | Yes | Yes | Yes | Yes |
| Total Messages for Mutual Auth. | 4-5$L$ | 4$L$ | 3$L$ | 4$L$ |
| Memory Size on Tag | 6$L$ | 7$L$ | 5$L$ | 7$L$ |
| Memory Size for each Tag on Database | 6$L$ | 4$L$ | 3$L$ | 4$L$ |
| Operation Types on Tag | $\oplus$, $\vee$, $\wedge$, $+$ | $\oplus$, $\vee$, $\wedge$, $+$, Rot | $\wedge$, $\vee$, $\oplus$, Rot | $\oplus$, $+$, Rot, $MixBits$ |

**3. Check of $C_1$.** If Step 2 ($\{2.0 - 2.2\}$) completely fails, the process is repeated from Step 1.

When messages $\{A_{n+1}, B_{n+1}\}$ are accepted by the genuine tag, the tag sends $C_{n+1}$ and updates its secret values. However, the reader, which is not involved in the attack, keeps on storing its old values. So the reader and the tag lose their synchronized state and this situation is irreversible.

The remaining question is to know how efficient the attack is. $C_1$ is restricted to having a hamming weigh of 2 in order for the hamming weigh of $N_n$ and $N_n \oplus C_1$ to be unknown but have a good probability of being equal. As two bits are flipped in $N_n$, and $N_n$ is a uniformly distributed random vector, the above condition is satisfied with a probability of $1/2$. Finally, the adversary has to test with different values of $C_2$. As the adversary does not know the hamming weight of $N_n \oplus C_1$, he can not say how many bits $C_1$ is rotated. However, he knows that the vector resulting from this rotation has a hamming weight of 2, which is quite advantageous. Indeed, the average number of times that the adversary has to try is $C_{L,2} = \binom{L}{2} = \binom{128}{2} = 8128 \ll 2^{128}$. ∎

Finally, a simple comparison of ultralightweight authentication protocols is shown in Table 1, where $L$ designates the bit length of variables used.

## 5  Conclusions

In this paper, we present the cryptanalysis of Lee et al. protocol, which is one of the most recent RFID mutual authentication protocols in the area of ultralightweight cryptography. The scheme presents noteworthy weaknesses related to most of the security properties initially required in its protocol design. Furthermore, the protocol is an excellent example as how triangular and non-triangular functions have to be combined to design secure ultralightweight protocols, and also about their combined usage does not guarantee any security by itself.

# References

[1] P. Peris-Lopez, J. C. Hernandez-Castro, J. M. Estevez-Tapiador, and A. Ribagorda. M2AP: A minimalist mutual-authentication protocol for low-cost RFID tags. In *Proc. of UIC'06*, volume 4159 of *LNCS*, pages 912–923. Springer-Verlag, 2006.

[2] P. Peris-Lopez, J. C. Hernandez-Castro, J. M. Estevez-Tapiador, and A. Ribagorda. LMAP: A real lightweight mutual authentication protocol for low-cost RFID tags. Hand. of Workshop on RFID and Lightweight Crypto, 2006.

[3] P. Peris-Lopez, J. C. Hernandez-Castro, J. M. Estevez-Tapiador, and A. Ribagorda. EMAP: An efficient mutual authentication protocol for low-cost RFID tags. In *Proc. of IS'06*, volume 4277 of *LNCS*, pages 352–361. Springer-Verlag, 2006.

[4] Hung-Yu Chien. "SASI: A New Ultralightweight RFID Authentication Protocol Providing Strong Authentication and Strong Integrity". *IEEE Transactions on Dependable and Secure Computing* 4(4):337–340. Oct.-Dec. 2007.

[5] Hung-Min Sun, Wei-Chih Ting, and King-Hang Wang. "On the Security of Chien's Ultralightweight RFID Authentication Protocol". Cryptology ePrint Archive. http://eprint.iacr.org/2008/083, 2008.

[6] Tianjie Cao, Elisa Bertino, and Hong Lei. "Security Analysis of the SASI Protocol". *IEEE Transactions on Dependable and Secure Computing*, 2008.

[7] P. D'Arco and A. De Santis. "From Weaknesses to Secret Disclosure in a Recent Ultra-Lightweight RFID Authentication Protocol". Cryptology ePrint Archive. http://eprint.iacr.org/2008/470, 2008.

[8] P. Peris-Lopez, J. C. Hernandez-Castro, J. M. Estevez-Tapiador, and A. Ribagorda. Advances in Ultralightweight Cryptography for Low-cost RFID Tags: Gossamer Protocol. In *Proc. of WISA'08*, Volume 5379 of *LNCS*, pages 56-68. Springer-Verlag, 2008.

[9] Y.-C. Lee, Y.-C. Hsieh, P.-S. You, T.-C. Chen A New Ultralightweight RFID Protocol with Mutual Authentication, In *Proc. of WASE'09*, Volume 2, of *ICIE*, pages 58-61, 2009.

[10] S. Weis. Security and Privacy in Radio-Frequency Identification Devices. In *Master Thesis, MIT*, 2003.

[11] T. Li and G. Wang. Security analysis of two ultra-lightweight RFID authentication protocols. In *Proc. of IFIP-SEC'07*, 2007.

[12] C. Hung-Yu and H. Chen-Wei. Security of ultra-lightweight RFID authentication protocols and its improvements. *SIGOPS Oper. Syst. Rev.*, 41(4):83–86, 2007.

[13] M. Bárász, B. Boros, P. Ligeti, K. Lója, and D. Nagy. "Breaking LMAP", *Proc. of RFIDSec'07*, 2007.

[14] M. Bárász, B. Boros, P. Ligeti, K. Lója, and D. Nagy. "Passive Attack Against the M2AP Mutual Authentication Protocol for RFID Tags", *Proc. of First International EURASIP Workshop on RFID Technology*, 2007.

[15] A. Klimov and A. Shamir. "New Applications of T-functions in Block Ciphers and Hash Functions". *Proc. of FSE'05*, LNCS vol. 3557, pp. 18–31. Springer-Verlag, 2005.

[16] R. Phan. Cryptanalysis of a new ultralightweight RFID authentication protocol - SASI. IEEE Transactions on Dependable and Secure Computing, 2008.

[17] J. C. Hernandez-Castro, J. M. E. Tapiador, P. Peris-Lopez, T. Li and J.-J. Quisquater. Cryptanalysis of the SASI Ultralightweight RFID Authentication Protocol with Modular Rotations. In *Proc. of WCC'09*, Lofthus, Norway, May 10-15, 2009.

[18] P. Peris-Lopez, J. C. Hernandez-Castro, J. M. Estevez-Tapiador, T. Li and J. C.A. van der Lubbe. Weaknesses in Two Recent Lightweight RFID Authentication Protocols. Hand. of Workshop on RFID Security, 2009.

[19] Juels, A., Weis, S.: Defining strong privacy for RFID. In: Proc. of PerCom 2007, IEEE Computer Society Press (2007) 342–347

[20] B. Alomair and R. Poovendran. On the Authentication of RFID Systems with Bitwise Operations. In: Proc. of NTMS'08, pages 1–6, 2008.

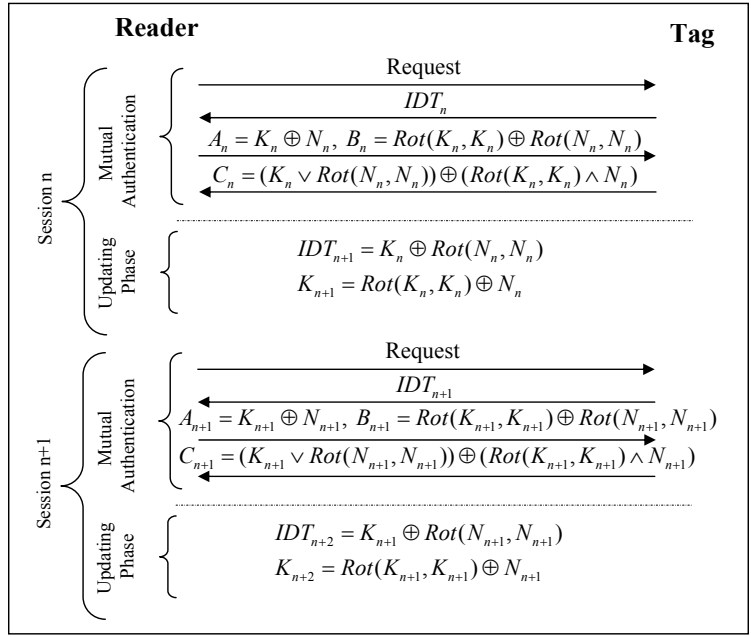

(a) Full Disclosure and Cloning Attacks

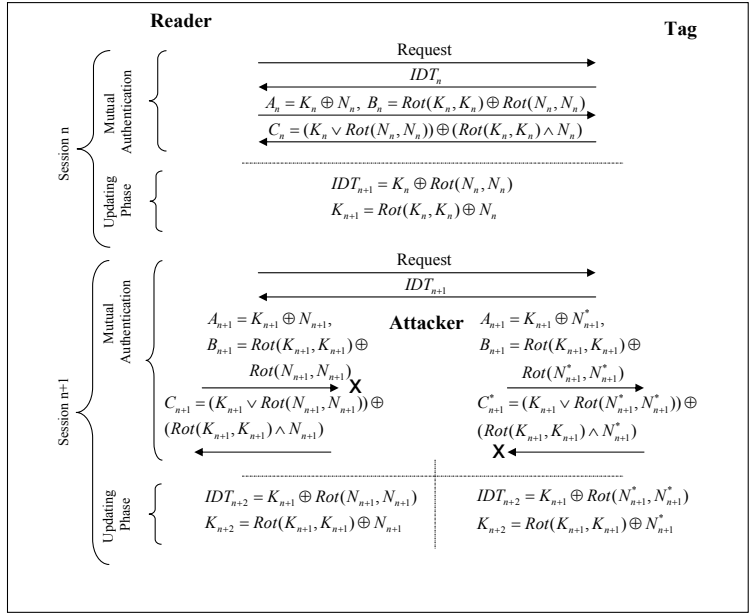

(b) De-synchronization Attacks

Figure 2: Passive and Active Attacks